\documentclass[12pt]{article}
\usepackage{epsfig}
\usepackage{graphicx}
\usepackage{amsmath}

\begin{document}
%%Define some new commands and macros
 \newcommand{\beq}{\begin{equation}}
\newcommand{\eeq}{\end{equation}}
\newcommand{\bea}{\begin{eqnarray}}
\newcommand{\eea}{\end{eqnarray}}
\newcommand{\beqn}{\begin{eqnarray}}
\newcommand{\eeqn}{\end{eqnarray}}
\newcommand{\beas}{\begin{eqnarray*}}
\newcommand{\eeas}{\end{eqnarray*}}
\newcommand{\defi}{\stackrel{\rm def}{=}}
\newcommand{\non}{\nonumber}
\newcommand{\bquo}{\begin{quote}}
\newcommand{\enqu}{\end{quote}}
\def\m12{\Delta m_{12}}
\def\m3{\Delta m_{13}}
\def\m24{\Delta m_{24}}
\newcommand{\ve}{\varepsilon}
%%%%%%%%%%%%%%%%%%%%%%%%%%%%%%%%%% definitions
%
\def\lsim{\mathrel{\rlap{\lower3pt\hbox{\hskip0pt$\sim$}}
    \raise1pt\hbox{$<$}}}
%less than or approx. symbol
%
\def\gsim{\mathrel{\rlap{\lower4pt\hbox{\hskip1pt$\sim$}}
    \raise1pt\hbox{$>$}}}
%greater than or approx. symbol
\def\de{\partial}
\def\Tr{ \hbox{\rm Tr}}
\def\const{\hbox {\rm const.}}
\def\o{\over}
\def\im{\hbox{\rm Im}}
\def\re{\hbox{\rm Re}}
\def\bra{\langle}\def\ket{\rangle}
\def\Arg{\hbox {\rm Arg}}
\def\Re{\hbox {\rm Re}}
\def\Im{\hbox {\rm Im}}
\def\diag{\hbox{\rm diag}}

\renewcommand{\theequation}{\thesection.\arabic{equation}}

\def\stroke{\vrule height8pt width0.4pt depth-0.1pt}
\def\topfleck{\vrule height8pt width0.5pt depth-5.9pt}
\def\botfleck{\vrule height2pt width0.5pt depth0.1pt}
\def\Zmath{\vcenter{\hbox{\numbers\rlap{\rlap{Z}\kern
0.8pt\topfleck}\kern
2.2pt\rlap Z\kern 6pt\botfleck\kern 1pt}}}
\def\Qmath{\vcenter{\hbox{\upright\rlap{\rlap{Q}\kern
3.8pt\stroke}\phantom{Q}}}}
\def\Nmath{\vcenter{\hbox{\upright\rlap{I}\kern 1.7pt N}}}
\def\Cmath{\vcenter{\hbox{\upright\rlap{\rlap{C}\kern
3.8pt\stroke}\phantom{C}}}}
\def\Rmath{\vcenter{\hbox{\upright\rlap{I}\kern 1.7pt R}}}
\def\Z{\ifmmode\Zmath\else$\Zmath$\fi}
\def\Q{\ifmmode\Qmath\else$\Qmath$\fi}
\def\N{\ifmmode\Nmath\else$\Nmath$\fi}
\def\C{\ifmmode\Cmath\else$\Cmath$\fi}
\def\R{\ifmmode\Rmath\else$\Rmath$\fi}

%%%%%%%%%%%%%%%%%%%%%%%%%%%%%%%%%%%%%%%%%%%%%%%%%%%%%%%%%%%%%%%%%%%%

\def\QATOPD#1#2#3#4{{#3 \atopwithdelims#1#2 #4}}
\def\stackunder#1#2{\mathrel{\mathop{#2}\limits_{#1}}}
\def\stackreb#1#2{\mathrel{\mathop{#2}\limits_{#1}}}
\def\Tr{{\rm Tr}}
\def\res{{\rm res}}
\def\Bf#1{\mbox{\boldmath $#1$}}
\def\balpha{{\Bf\alpha}}
\def\bbeta{{\Bf\beta}}
\def\bgamma{{\Bf\gamma}}
\def\bnu{{\Bf\nu}}
\def\bmu{{\Bf\mu}}
\def\bphi{{\Bf\phi}}
\def\bPhi{{\Bf\Phi}}
\def\bomega{{\Bf\omega}}
\def\blambda{{\Bf\lambda}}
\def\brho{{\Bf\rho}}
\def\bsigma{{\bfit\sigma}}
\def\bxi{{\Bf\xi}}
\def\bbeta{{\Bf\eta}}
\def\d{\partial}
\def\der#1#2{\frac{\d{#1}}{\d{#2}}}
\def\Im{{\rm Im}}
\def\Re{{\rm Re}}
\def\rank{{\rm rank}}
\def\diag{{\rm diag}}
\def\2{{1\over 2}}
\def\ntwo{${\cal N}=2\;$}
\def\4N{${\cal N}=4$}
\def\none{${\cal N}=1\;$}
\def\x{\stackrel{\otimes}{,}}
\def\ba{\beq\new\begin{array}{c}}
\def\ea{\end{array}\eeq}
\def\be{\ba}
\def\ee{\ea}
\def\stackreb#1#2{\mathrel{\mathop{#2}\limits_{#1}}}

\def\Tr{{\rm Tr}}
\newcommand{\vp}{\varphi}
\newcommand{\pt}{\partial}

\setcounter{footnote}0
%\begin{titlepage}
%\renewcommand{\thefootnote}{\fnsymbol{footnote}}

\vfill

%%%%%%%%%%%%%%%%%%%%%%%%%%%%%%%%
\begin{titlepage}

\begin{flushright}
FTPI-MINN-06/11, UMN-TH-2439/06\\
ITEP-TH-10/06 \\
April 27, 2006
\end{flushright}

\begin{center}

{ \Large \bf Bulk--Brane Duality in Field Theory
    }
\end{center}

\begin{center}
 {
 \bf    M.~Shifman$^{a}$ and \bf A.~Yung$^{a,b,c}$}
\end {center}
\vspace{0.3cm}
\begin{center}

$^a${\it  William I. Fine Theoretical Physics Institute,
University of Minnesota,
Minneapolis, MN 55455, USA}\\
$^{b}${\it Petersburg Nuclear Physics Institute, Gatchina, St. Petersburg
188300, Russia\\
$^c${\it Institute of Theoretical and Experimental Physics, Moscow
117259, Russia}}
\end{center}

\vspace*{.45cm}
\begin{center}
{\large\bf Abstract}
\end{center}

We consider (3+1)-dimensional \ntwo supersymmetric QED with two flavors
of fundamental hypermultiplets. This theory supports 1/2-BPS domain walls and flux
tubes (strings), as well as their 1/4-BPS junctions. The effective
(2+1)-dimensional theory on the domain wall is known to be a U(1) gauge
theory. Previously, the wall--string junctions were shown
to play the role of massive charges in this theory.  However, the field theory
of the junctions on the wall (for semi--infinite strings) appears to be inconsistent
due to infrared problems.  All these problems can be eliminated by
compactifying one spatial dimension orthogonal to the wall and considering a
wall--antiwall system ($W\bar W$) on a {\sl cylinder}. We argue that for
certain values of parameters this set-up provides an example of a
controllable analog of
bulk--brane duality in field theory. Dynamics of the 4D bulk are mapped onto 3D
boundary theory: 3D \ntwo SQED with two matter superfields  and a weak--strong
coupling constant relation in 4D and 3D, respectively. The cylinder radius is seen as a ``real
mass" in 3D \ntwo SQED. We work out (at weak coupling) the quantum version of the
world-volume theory on the walls. Integrating out massive
matter (strings in the bulk theory) one generates a Chern--Simons term on the wall
world volume and an interaction between the wall and antiwall that scales as
a {\sl power} of distance. Vector and scalar (classically) massless excitations on the walls
develop a mass gap at the quantum level; the long-range interactions disappear.
The above duality implies that the wall and its antiwall partner (at strong coupling
in the bulk theory) are stabilized at the opposite sides of the cylinder.

\vspace*{.05cm}

\end{titlepage}

\section{Introduction}
\label{intro}
\setcounter{equation}{0}

Duality is one of the most fascinating ideas in
theoretical physics. Basically,  duality means that one and the same dynamical
system can be described using two different theories. Moreover, when
one description is in the weak coupling regime, the second is at strong coupling
and {\em vice versa}. This makes duality extremely hard to prove.
Exceptions are theories where one can find exact
solutions and continue them from weak to strong coupling. The
most well-known example is electromagnetic duality in \ntwo gauge
theories \cite{SW1,SW2}. In other cases duality is conjectured, e.g. AdS/CFT correspondence \cite{Mal,GuPol,ew98}, and then indirectly verified.
On the other hand, once duality is confirmed, it becomes a
powerful tool to study theories at strong coupling using their
weakly coupled duals.

AdS/CFT correspondence provides an example of duality between theories
with different dimensions. On the one side is a string theory in the bulk
which, in a certain limit, reduces to a classical supergravity model.
A gauge theory on the four-dimensional boundary of the five-dimensional
bulk is on the other side of AdS/CFT.

In this paper we present a purely field-theoretic example of such ``holographic"
duality. We  consider  (3+1)-dimensional \ntwo gauge theory with domain
walls and flux tubes (strings) as our bulk model. One of the spatial dimensions is compactified, so that in fact our four-dimensional bulk is $R^3\times S^1$.
We show that a system of two parallel walls $W\bar W$ and strings stretched between these walls has a dual description in terms of (2+1)-dimensional  U(1) gauge theory
on the world volume of domain walls. In three dimensions the dual model has
${\cal N}=2$ (four supercharges). Moreover, the weak coupling regime
in the bulk theory corresponds to strong coupling in the world-volume
theory and {\em vice versa}. As it often happens, we cannot rigorously
prove duality in this case. However, we are able to provide indirect evidence and to  explicitly  determine a condition which ensures duality.

Our bulk theory is \ntwo supersymmetric QED with 2 flavors of charged
matter hypermultiplets and the Fayet-Iliopoulos term \cite{FI}. In this theory one can construct both,
the 1/2 BPS-saturated walls and strings. Moreover, the 1/4 BPS-saturated wall-string
junctions exist too \cite{SYw}. In the domain wall world-volume theory these
junctions are seen as ``electric charges" of dual QED.
However, construction of the world-volume theory, with
the electric charges (a.k.a. junctions) {\sl included}, was hampered by the
fact that the string attached to the given junction was assumed to be semi-infinite
\cite{SYw}. Thus, it could not be mapped to any local theory
on the domain wall. A way out was suggested by Tong
in the inspiring paper \cite{Tbrane}.
He suggested to consider two parallel domain walls
at finite distance from each other, connected
by a finite-length string. As will be discussed in the bulk of the paper,
the concrete set-up  presented in \cite{Tbrane} has no mass shell,
due to long-range (scalar-field-mediated) interactions
which result in an infinite bending of the walls at infinity.
Building on Tong's proposal and our previous results we develop
a consistent construction implementing the bulk--brane duality in field theory.
To the best of our knowledge, this is the first example of this type.
We hasten to warn the reader that the above duality must be understood in a limited sense. Usually
the phrases ``brane-bulk duality" or ``holographic description"   apply
to AdS/CFT where the the boundary theory sees {\em everything} that
happens in the bulk. This certainly is not true of the theory on
the domain walls we discuss. For example, our 3D theory
 has no way of knowing about such
degrees of freedom in the bulk as particles collisions
far from the branes. In this sense, we deal with a ``blurred holography."

We show that infrared problems can be eliminated by
compactifying one spatial dimension orthogonal
to the wall and considering a wall-antiwall system on a cylinder.
If the radius of the cylinder is much larger than the wall thickness,
classically, the relative distance between the walls is a modulus.
The corresponding excitations are gapless. Supersymmetry then predicts
that other states from the same supermultiplet, namely, the photon and its fermion
partners, are massless too. If we place classical static sources of the dual electromagnetic field and the scalar field
on the wall, the induced
interactions  do not fall off at infinity. A generic string--brane configuration
has infinite energy. A finite energy is obtained on configurations
of a very special type --- with equal number of string connections on
both sides of the cylinder.

We work out the quantum version of the 3D theory on the walls and show that
taking account of quantum effects  drastically changes the situation.
The massive matter
(a.k.a. strings in the bulk theory) generates a Chern--Simons term on the wall world volume and produces a quadratically rising interaction potential
between the wall and antiwall, with the minimum
of the potential energy  corresponding to the
wall and antiwall located at the opposite sides of the cylinder.
The moduli of the classical theory are lifted. There are no long-range forces on the
wall.

The paper is organized as follows. In Sect. \ref{bwv} we introduce our bulk
theory and briefly review  BPS domain wall and string solutions in this
theory. In Sect. 3 we review the 1/4-BPS solution for the string-wall junction
and represent this junction as a classical charge in the (2+1)-dimensional
low-energy effective theory on the wall. In Sect. \ref{quant} we
consider wall-antiwall system on a cylinder and suggest a quantum version
of the theory on the walls. We also discuss bulk--brane duality.
In Sect. \ref{physi} we discuss physics of the  world-volume theory and
show that both the Chern--Simons term and quadratic wall-antiwall interaction
potential are generated at the quantum level. Section 6 contains our conclusions
while Appendix deals with formulation of (2+1)-dimensional QED and parity
anomaly.

\section{Domain walls and strings}
\label{bwv}
\setcounter{equation}{0}

The bulk theory which we will work with
is \ntwo SQED with 2 flavors. It supports  both, the BPS-saturated domain walls and, if the Fayet--Iliopoulos term is added, the BPS-saturated ANO strings.
The bosonic part of the action of the bulk theory is\,\footnote{Here we use Euclidean notation; we switch to Minkowskian notation in Sect.~\ref{physi} and Appendix.}
\beqn
&& S=\int d^4 x \left\{ \frac{1}{4 g^2} F_{\mu \nu}^2 +
\frac{1}{g^2} |\partial_\mu a|^2 +\bar{\nabla}_\mu \bar{q}_A \nabla_\mu q^A +
\bar{\nabla}_\mu \tilde{q}_A \nabla_\mu \bar{\tilde{q}}^A
\right.\nonumber\\[3mm]
 &&
 +
 \left.\frac{g^2}{8}\left(|q^A|^2-|\tilde{q}_A|^2-\xi\right)+\frac{g^2}{2}
\left|\tilde{q}_A q^A\right|^2
+\frac{1}{2}(|q^A|^2+|\tilde{q}^A|^2)\left| a+\sqrt{2}m_A\right|^2 \right\}, \nonumber\\
\label{n2sqed}
\eeqn
where
\beq
\nabla_\mu=\partial_\mu-\frac{i}{2}A_\mu\,,\qquad
\bar{\nabla}_\mu=\partial_\mu+\frac{i}{2}A_\mu\,.
\eeq
Here $\xi$ is the coefficient in front of the Fayet-Iliopoulos term, $g$ is the U(1)
gauge coupling, the
index $A=1,2$ is the flavor index; and the mass parameters $m_1,m_2$ are
assumed to be  real. In addition we will assume
\beq
\Delta m \equiv
m_1-m_2\gg g\sqrt{\xi} \,.
\label{mxi}
\eeq
Simultaneously, $\Delta m \ll  (m_1+m_2)/2$.
There are two vacua in this theory: in the first vacuum
\beq
a=-\sqrt{2} m_1,\qquad q_1=\sqrt{\xi}, \qquad  q_2=0\,,
\label{fv}
\eeq
and in the second one
\beq
a=-\sqrt{2} m_2, \qquad q_1=0,  \qquad q_2=\sqrt{\xi}\,.
\label{sv}
\eeq
The vacuum expectation value (VEV) of the
field $\tilde{q}$ vanishes in both  vacua.
Hereafter we will stick to the {\em ansatz} $\tilde{q}=0$.

\begin{figure}[h]
%\epsfxsize=9cm
%\epsfysize=2.5cm
%\centerline{\epsfbox{syfig3}}
%\centerline{\epsfbox{syfig3.eps}}
 \centerline{\includegraphics[width=3.5in]{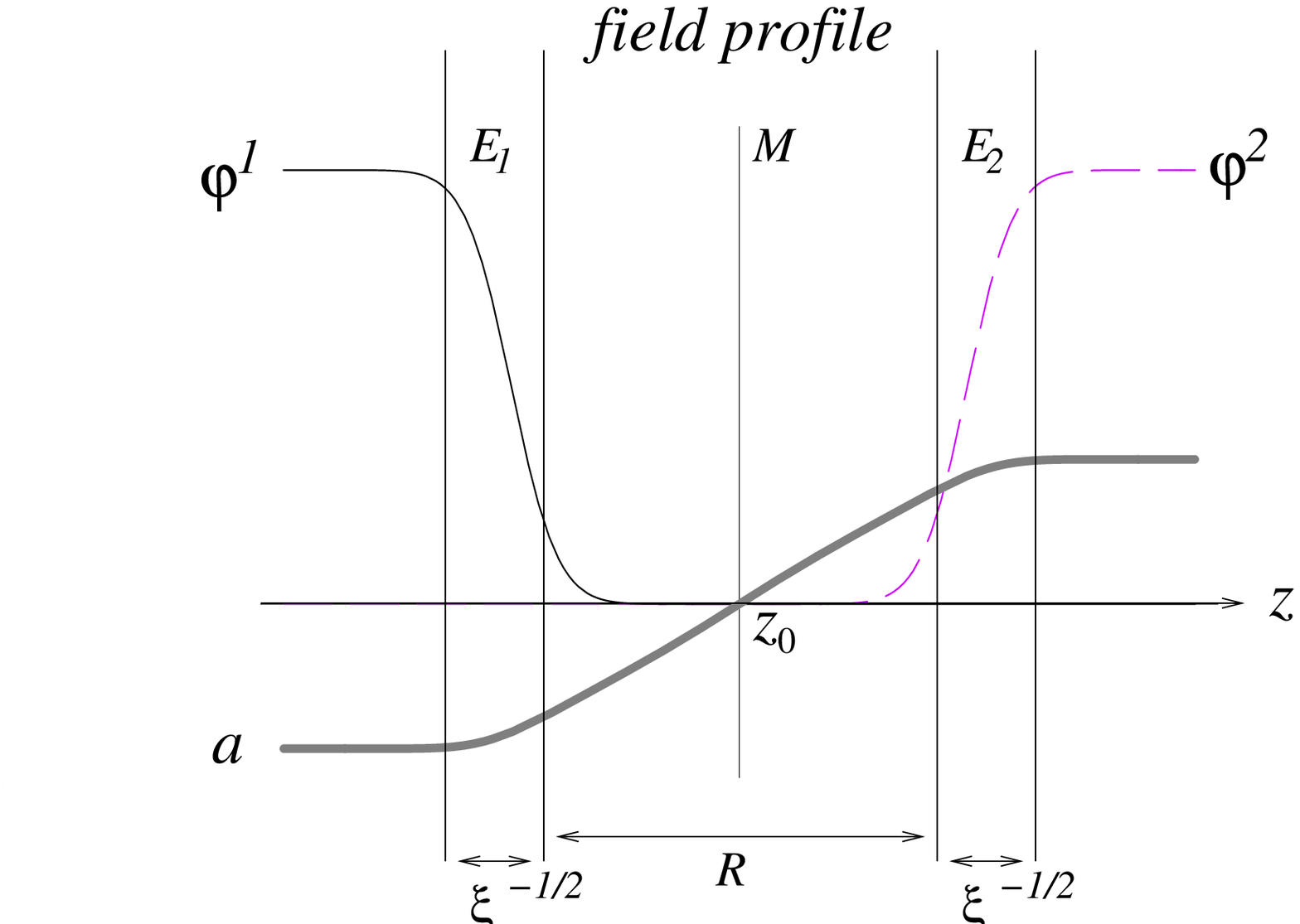}}
 \caption{\footnotesize Internal structure of the domain wall:
two edges (domains $E_{1,2}$) of the width $\sim \xi^{-1/2}$
are separated by a broad middle band (domain $M$) of the width $R
\sim \Delta m /(g^2\xi)$.}
\label{syfigthree}
\end{figure}

A BPS domain wall
interpolating between the two vacua of our bulk
theory was explicitly constructed in Ref. \cite{SYw}.
Assuming that all fields depend only on the coordinate
$z=x_3$, it is possible to write the energy in the Bogomol'nyi form
\cite{Bogomolny},
\beqn
  E &=&
 \int dx_3 \left\{ \left|\nabla_3 q^A \pm \frac{1}{\sqrt{2}}q^A
 (a+\sqrt{2}m_A)\right|^2\right.
 \nonumber\\[3mm]
&+&\left. \left|\frac{1}{g}\partial_3 a \pm \frac{g}{2 \sqrt{2}}
\left(|q^A|^2-\xi\right)\right|^2
\pm \frac{1}{\sqrt{2}} \xi \partial_3 a\right\}.
\label{bogfw}
\eeqn
Requiring the first two terms above to vanish gives us the BPS equations
for the wall. Assuming that $\Delta m >0$ we choose the upper sign in
(\ref{bogfw}). The tension is given by the total derivative term
(the last one in Eq.~(\ref{bogfw}))
which can be identified as the $(1,0)$ central charge
of the supersymmetry algebra,
\beq
T_{\rm w}=\xi \, \Delta m\,.
\label{wten}
\eeq
The wall solution has a three-layer structure
(see Fig.~\ref{syfigthree}):
in the two outer layers (which have width
${O}(({g \sqrt{\xi}})^{-1}$)) the squark fields drop
to zero exponentially;  in the inner layer
the field $a$ interpolates between its two vacuum values.
The  thickness of this inner layer is given by
\beq
\label{R}
R=\frac{4 \Delta m}{g^2 \xi}\,.
\\[4mm]
\eeq

This wall is an $1/2$ BPS solution of the Bogomol'nyi equations.
In other words, the soliton breaks four of eight supersymmetry
generators of the \ntwo bulk theory.
In fact, as was shown in \cite{SYw}, the four supercharges selected by
the conditions
\beqn
\bar{\ve}^2_{\dot{2}}&=& -i\ve^{21}\,,\qquad
\bar{\ve}^1_{\dot{2}}=-i\ve^{22}\,,
\nonumber\\[3mm]
\bar{\ve}^1_{\dot{1}}&=& i\ve^{12}\, ,\qquad
\bar{\ve}^2_{\dot{1}}=i\ve^{11}\,,
\label{wsusy}
\eeqn
act trivially on the wall solution. Here $\ve^{\alpha f}$ and
$\bar{\ve}^f_{\dot{\alpha}}$ are eight supertransformation parameters.

The moduli space
is described by two bosonic coordinates:
one of these coordinates is associated with the wall translation;
the other one is a U(1)  compact parameter $\sigma$. Its origin
is as follows \cite{SYw}.
The bulk theory at $\Delta m\neq 0$ has U(1)$\times$U(1) flavor symmetry
corresponding to two independent rotations of two quark flavors. In both
vacua only one quark develops a  VEV. Therefore, in both vacua only one of these
two U(1)'s is broken. The corresponding phase is eaten by the Higgs mechanism.
However, on the wall both quarks have nonvanishing values, breaking both U(1)
groups. Only one of corresponding  two phases is eaten by the Higgs mechanism.
The other one becomes a Goldstone mode living on the wall.

It is possible to promote these   moduli
to fields depending on the wall coordinates $x_n$ ($n=0,1,2$). Then, deriving
a world-volume theory for the moduli fields on the wall
is straightforward. The U(1) phase $\sigma$
discussed above can be dualized \cite{apolya} to a
U(1) gauge field in (2+1) dimensions. Thus, the world-volume theory is
a U(1) gauge theory. The domain wall
under consideration can be interpreted as a $D$-brane prototype
in field theory \cite{2,sdw,GPTT,SYw}. The bosonic part of the world-volume
action is
\beq
S_{2+1}=\int d^3 x \left\{ \frac{T_{\rm w}}{2}
\left(\partial_n z_0\right)^2-\frac{1}{4 e^2} \left(
F_{mn}^{(2+1)} \right)^2   \right\}
,
\label{3wwa}
\eeq
where $z_0$ describes the translational mode while
$$e^2\equiv e^2_{2+1}$$ is the coupling constant of the
effective  U(1) theory on the wall related to the parameters of the bulk theory as
\beq
e^2=4 \pi^2 \frac{\xi}{\Delta m}\,.
\label{3cc}
\eeq
The fermion content of
the world-volume theory
is given by two three-dimensional Majorana spinors, as
is required by ${\cal N}=2$ in three dimensions.
The full world-volume theory is a
 U(1) gauge theory in $(2+1)$ dimensions, with
four supercharges. The Lagrangian and the
corresponding superalgebra
can be obtained by reducing four-dimensional
$\mathcal{N}=1$ SQED (with no matter)
to three dimensions (see Appendix).

The field $z_0$ in (\ref{3wwa}) is the \ntwo superpartner of the gauge field
$A_n$. To make it more transparent we make a rescaling, introducing a new field
\beq
a_{2+1}=2\pi\xi\,z_0\,.
\label{az}
\eeq
In terms of $a_{2+1}$ the action (\ref{3wwa}) takes the form
\beq
S_{2+1}=\int d^3 x \left\{ \frac{1}{2e^2}
\left(\partial_n a_{2+1}\right)^2-\frac{1}{4 e^2} \left( F_{mn}^{(2+1)} \right)^2   \right\}.
\label{pure}
\eeq
The gauge coupling constant $e^2$ has dimension of mass in three
dimensions. A characteristic scale of massive excitations on the world
volume theory is of the order of the inverse thickness of the wall $1/R$,
see (\ref{R}). Thus the dimensionless parameter that characterizes the
coupling strength in the world-volume theory is $e^2 R$,
\beq
e^2 R=\frac{16\pi^2}{g^2}.
\label{duality}
\eeq
We interpret this as a feature of the bulk--wall duality:
the weak coupling regime in the bulk theory
corresponds to strong coupling on the wall and {\em vice versa} \cite{SYw}.
Of course, finding explicit domain wall solutions and deriving the effective
theory on the wall assumes weak coupling regime in the bulk, $g^2\ll 1$.
In this limit the world-volume theory is in the strong coupling regime
and is not very useful.

Our theory (\ref{n2sqed})  is Abelian and as such does not have
monopoles. However,  we could have compactified U(1),
starting from the \ntwo non-Abelian  theory with
the gauge group SU(2), broken down to U(1)
by condensation of the adjoint scalar field (whose third component
is $a$, see (\ref{fv}) and (\ref{sv})). This non-Abelian theory would
have the 't Hooft-Polyakov monopoles \cite{thopo}. In the low-energy limit
(\ref{n2sqed}) these monopoles
become heavy external magnetic charges, with mass of the
order of $m_{1,2}/g^2$.
Once electrically charged fields condense in both vacua (\ref{fv})
and (\ref{sv}), these monopoles are in the confining phase in both
vacua.

In fact, in each of the two vacua of the bulk theory the
magnetic charges are confined by the Abrikosov--Nielsen--Olesen
(ANO) strings \cite{ANO}.
In the vacuum in which $a=-\sqrt{2} m_1$ we can use the
{\em ansatz} $q_2=0$ and
write the following BPS equations:
\beq
F_3^*-\frac{g^2}{2}\left(|q^1|^2-\xi\right)=0\,, \qquad
\left(\nabla_1-i\nabla_2\right)q^1=0,
\label{fluxone}
\eeq
where
$$
F^{*}_i=\frac{1}{2}\, \varepsilon_{ijk} F_{jk}\,,\qquad i,j,k=1,2,3\,.
$$
Analogous equations can be written for the flux tube
in the other vacuum, $a=-\sqrt{2} m_2$.
These objects are half-critical, much in the same way as the domain walls above.

The magnetic flux of the minimal-winding flux tube
is $4 \pi$, while its    tension   is given by
the $(1/2,1/2)$ central charge \cite{GS},
\beq
T_{\rm s} =2 \pi \xi\,.
\label{tens}
\eeq
The thickness of the tube is of the order of
\beq
(g^2\,\xi)^{-1/2}\,.
\label{ssize}
\eeq

\section{Wall--string junctions}
\label{vwj}
\setcounter{equation}{0}

Let us consider  a flux tube ending on a domain wall.
The flux tube is semi-infinite and aligned perpendicular to the wall.
This configuration was studied in gauge theories in Refs.~\cite{SYw,SYnawall,J14,SakTong}
(for a review see \cite{Trev,Jrev})  and, earlier,
 in four-dimensional sigma models \cite{GPTT}.

Consider a vortex oriented in the $z$ direction
ending on a wall oriented in the $(x_1,\, x_2)$ plane;
let the string extend at $z>0$ and let the magnetic
flux be oriented in the negative $z$ direction.
The BPS first-order equations can be written for the composite soliton  \cite{SYw},
\beqn
&&
 F_1^*-iF_2^* - \sqrt{2}\left(\partial_1-i\partial_2\right) a=0\,,
 \nonumber\\[2mm]
&&
F_3^*-\frac{g^2}{2} \left(|q^A|^2 - \xi\right)-\sqrt{2} \partial_3 a=0\,,
\nonumber\\[2mm]
&&
\nabla_3 q^A=-\frac{1}{\sqrt{2}}q^A \left(a+\sqrt{2}m_A\right),
\nonumber\\[2mm]
&& \left(\nabla_1-i\nabla_2\right) q^A=0\,.
 \label{bpseq}
 \eeqn
These equations generalize both the $1/2$-BPS wall and string equations
and were used to determine the general $1/4$-BPS solution for the
string-wall junction, (i.e. a flux tube ending on the wall).

These equations were derived in \cite{SYw} as follows. The first-order
equations for 1/2-BPS string can be obtained by imposing
the requirement that four
supercharges of the bulk theory selected by the conditions \cite{VY}
\beqn
\ve^{12}=-\ve^{11}\,,\qquad
\bar{\ve}^2_{\dot{1}}=-\bar{\ve}^1_{\dot{1}}\,,
\nonumber\\[3mm]
\ve^{21}=\ve^{22}\,,\qquad
\bar{\ve}^1_{\dot{2}}=-\bar{\ve}^2_{\dot{2}}\,,
\label{ssusy}
\eeqn
act trivially on the string solution. Then, imposing both wall and
string conditions (\ref{wsusy}) and (\ref{ssusy}) to select two
supercharges
which act trivially on the string-wall junction, we obtain the
first-order equations (\ref{bpseq}).

The solution for the string-wall junction
at large distance $r$ from the string end-point has the form of a wall solution
with the collective coordinates $z_0$ and the U(1) phase $\sigma$
depending on the
world-volume coordinates $x_1$ and $x_2$. Namely \cite{SYw}, the
 wall is logarithmically
bent due to the fact that the vortex pulls it,
\beq
z_0= -\frac{1}{\Delta m} \ln r+{\rm const}\, .
\label{zbend}
\eeq
Moreover, the magnetic flux from the string penetrates into the wall
and, therefore, the string end-point is seen as an electric charge in the
world-volume theory, dual QED.\footnote{Here the word ``dual" is used in the sense of
electromagnetic (nonholographic) duality, with the phase $\sigma$
dualized \`a la Polyakov \cite{apolya} to be traded for the 3D gauge field $A_n$.}
The electric field
at large distances $r$ from the string-wall junction
is given by
\beq
F_{0i}^{2+1}=\frac{e^2}{2 \pi } \frac{x_i}{r^2} \,,
\label{elf}
 \eeq
where $i=1,2$.

In the world-volume theory {\em per se}, the fields (\ref{zbend})
and (\ref{elf}) can be
considered as produced by classical point-like charges which interact in a
standard way with the electromagnetic field $A_n$ and the scalar field
$a_{2+1}$,
\beqn
S_{2+1}&=&\int d^3 x \left\{ \frac{1}{2e^2}
(\partial_n a_{2+1})^2-\frac{1}{4 e^2} ( F_{mn}^{(2+1)} )^2
\right.
\nonumber\\[4mm]
&+& A_n\,j_n - a_{2+1}\,\rho\Big\},
\label{cl}
\eeqn
where the classical electromagnetic current and the charge density of a
static  charge is given by
\beq
j_n (x)=\left\{ \delta^3 (x),\, 0,\, 0\right\}\,,\qquad \rho(x)=\delta^3 (x)
\label{classcur}
\eeq
for a unit charge located at $x_n =0$.
To derive (\ref{cl}) we used (\ref{zbend}) and (\ref{elf}),
as well as the relation (\ref{az}). Thus, the string is seen in the
world-volume theory on the wall as a unit static charge.

It is easy to calculate the energy of this static charge.
There are two distinct contributions to this energy \cite{ASYb}.
The first contribution is due to the gauge field,
\beqn
E^G_{(2+1)} &=& \int
\frac{1}{2 e^2_{2+1}} (F_{0i})^2\,  2 \pi r\, dr
\nonumber\\[4mm]
&=&\frac{\pi \xi}{\Delta m
}\int  \frac{ dr}{r}=\frac{\pi \xi}{\Delta m
} \ln {\left(g\sqrt{\xi}L\right)}.
\label{gfcc}
\eeqn
The integral $\int dr/r$ is logarithmically divergent both in the ultraviolet and
infrared. It is clear that the UV divergence is cut off at the transverse size of the string
(\ref{ssize}) and presents no problem. However, the infrared divergence is
much more serious. We introduced a large size $L$ to regularize it in
(\ref{gfcc}).

The second contribution, due to the $z_0$  field, is proportional to $\int dr/r$
too,
\beqn
E^H_{(2+1)} &=& \int_{r_0}^{r_f} \, \frac{T_{\rm w}}{2} \left(\partial_r
\, z_0 \right)^2
\, 2 \pi r\,  dr
\nonumber\\[4mm]
&=&
\frac{\pi \xi}{\Delta m
}\ln{\left(g\sqrt{\xi}L\right)}\,.
\label{zfcc}
\eeqn
Both contributions are logarithmically divergent in the infrared.
Their  occurrence is an obvious
feature of charged objects coupled to massless fields
in $(2+1)$ dimensions due to the
fact that the fields $A_n$ and $a_{2+1}$
do not die off at infinity which means infinite energy.

The above two contributions are equal (with the logarithmic
accuracy), even though their physical
interpretation is different. The total  energy of the string junction is
\beq
E^{G+H}=\frac{2\pi \xi}{\Delta m
}\ln{\left(g\sqrt{\xi}L\right)}\,.
\label{eto}
\eeq

We see that in our attempt to include  strings  as point-like
charges in the world-volume theory (\ref{cl}) we encounter problems
 already at the classical level. The energy of a single charge is
IR divergent.
(This is on top of the fact that semi-infinite strings in the bulk
have masses proportional to their length; once the length is infinite so are
their masses. For the time being we will disregard the latter aspect, to be addressed below.)

It is clear that the infrared problems will become even more severe in
quantum theory. One might suggest to overcome this problem by adding
antistrings in our picture. The antistring carries an  opposite magnetic flux
to that of the string; therefore, it  produces the following  (dual)
electric field on the wall:
\beq
F_{0i}^{2+1}=-\frac{e^2}{2 \pi }\, \frac{x_i}{r^2} \,,
\label{elfanti}
 \eeq
while the bending of the wall due to the string stretched at $z>0$ is still given
by Eq.~(\ref{zbend}). The fields (\ref{zbend}), (\ref{elf}) of  the string
and (\ref{zbend}), (\ref{elfanti}) of the antistring can be
described by the world-volume action (\ref{cl}) where the currents $j_n$ and
$\rho$  are given by
\beq
j_n (x)=n_e\{ \delta^3 (x),0,0\}\,,\qquad\rho(x)=n_s\delta^3 (x)\,.
\label{clcur}
\eeq
Here $n_e$ and $n_s$ are electric and scalar charges
 associated with the string end
with respect to the electromagnetic field $A_n$ and the scalar field $a$, respectively,
 $$ n_{e,s}=\pm 1\,.$$
The string with incoming flux has the charges
$(n_e,n_s)=(+1,+1)$
 while the string with outgoing flux (the antistring)
the charges $(n_e,n_s)=(-1,+1)$.

\begin{figure}[h]
%\epsfxsize=10cm
%\epsfxsize=6cm
%\epsfysize=2.5cm
%\centerline{\epsfbox{sbars.eps}}
 \centerline{\includegraphics[width=2in]{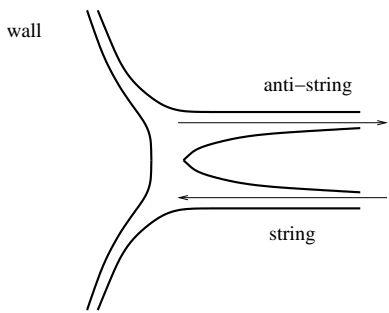}}
 \caption{\footnotesize String and anti-string ending on the wall. Arrows denote
the direction of the magnetic flux.
}
\label{fig:sbars}
\end{figure}

Now we can consider a configuration with equal number of strings and
antistrings  so that at large distances from their end-points the electric
field $A$ has a power fall-off and  produces no IR divergence,
see Fig.~\ref{fig:sbars}. This solves only a
half of the problem, however. The electric fields produced by the string and
antistring   cancel each other  at large $r$. At the same time, the bending of the wall is
doubled, as it is clear from Eq.~(\ref{clcur}) and Fig.~\ref{fig:sbars}.
Thus, the wall bending energy is still IR divergent.\footnote{This aspect was
 omitted in
\cite{Tbrane}. For further discussion see Sect.~\ref{quant}.}

A way out was suggested in \cite{ASYb}. For the infrared divergences to cancel
we should consider strings
coming to the wall from the right {\sl and} from the left. Clearly, the bending
of the wall produced by the string coming from the left is given by
\beq
z_0= \frac{1}{\Delta m} \ln r+{\rm const}\, .
\label{zbendl}
\eeq
It has the opposite sign as compared to the bending of  the string coming
from the right.
Thus, we have the following set of the electric (scalar) charges
associated with the  string end-points:
\beqn
&&
n_e=+1,\;\;\; \mbox{incoming flux},
\nonumber\\[2mm]
&&
n_e=-1,\;\;\; \mbox{outgoing flux},
\label{elch}
\eeqn
while their   scalar charges are
\beqn
&&
n_s=+1,\;\;\; \mbox{string from the right},
\nonumber\\[2mm]
&&
n_s=-1,\;\;\; \mbox{string from the  left}.
\label{scch}
\eeqn

\begin{figure}[h]
%\epsfxsize=10cm
%\epsfxsize=6cm
%\epsfysize=2.5cm
%\centerline{\epsfbox{strtstr.eps}}
 \centerline{\includegraphics[width=2in]{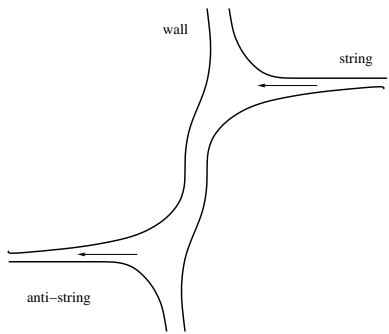}}
 \caption{\footnotesize String and anti-string ending on the wall
from different sides. Arrows denote
the direction of the magnetic flux.
}
\label{fig:strtstr}
\end{figure}

Clearly, the bending of the wall produced by two  string attached from
different sides of the wall tends to zero at large $r$ from the string end-points
and produces no IR divergence. In fact, it was shown in \cite{ASYb}
that the configuration depicted in Fig.~\ref{fig:strtstr} is a
non-interacting 1/4-BPS configuration. All logarithmic contributions are canceled; the
junction energy in this geometry is given by a finite negative contribution
\beq
E=-\frac{8\pi}{g^2}\,\Delta m\,,
\label{boojum}
\eeq
which is called the {\sl boojum energy} \cite{SakTong,ASYb}.

 In Sect.~\ref{quant}  we will consider a wall configuration in which the
attached strings
can appear from both sides of the wall. Then we will
suggest a quantum version of the theory on the wall.

\section{Quantizing strings on the wall}
\label{quant}
\setcounter{equation}{0}

In this section we will work out a quantum version of the world-volume theory
(\ref{cl}) with charged matter fields which, on the world volume,
represent strings of the bulk theory.
First we proceed in the spirit of Ref.~\cite{Tbrane}, and then deviate in a bid for
a theory free of IR divergences. A novel element is introduction of
two types of charged matter, to represent both, strings attached
from the left and from the right of the wall.

The mass of the string is equal to its tension (\ref{tens}) times its
length. If we have a single wall, all strings attached to it have infinite
length; therefore, they are infinitely heavy.  In the world-volume
theory on the wall (\ref{cl}) they are seen as classical infinitely heavy point-like charges.
In order
to quantize  these charges one has to make their masses finite. To this end
one  needs at least two domain walls \cite{Tbrane}.

\begin{figure}[h]
%\epsfxsize=10cm
%\epsfxsize=6cm
%\epsfysize=2.5cm
%\centerline{\epsfbox{cylinder.eps}}
\centerline{\includegraphics[width=2in]{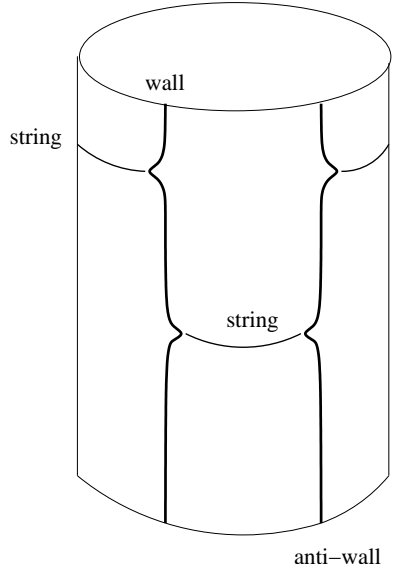}}
 \caption{\footnotesize A wall and antiwall connected by strings on
the cylinder. The circumference of the circle (the transverse slice of the cylinder)
is $L$.
}
\label{fig:cylinder}
\end{figure}

Let us describe our set-up in some detail.
First, we compactify the $x_3=z$ direction in our bulk theory (\ref{n2sqed}),
on a circle of length  $L$. Then we consider a pair ``wall plus antiwall" oriented
in the $\{x_1,\,x_2\}$ plane,  separated by a distance $l$ in the perpendicular direction,
see Fig.~\ref{fig:cylinder}. The wall and antiwall
experience attractive forces. Strictly speaking,
 this is not a BPS configuration ---
supersymmetry in the world-volume theory is broken. However, the wall-antiwall
interaction due to overlap of their profile functions is
exponentially suppressed at large separations,
\beq
L\sim l\gg R\,,
\label{largesep}
\eeq
where $R$ is the wall size (see Eq.~(\ref{R})). In what follows we {\em neglect
exponentially suppressed effects}. If so,  we neglect effects which
break  supersymmetry in our (2+1)-dimensional world-volume
theory. Thus, it continues to have
 four conserved supercharges (\ntwo supersymmetry
in (2+1) dimensions) as was the case for the isolated single wall.

Although our set-up contains the wall-antiwall pair,
in fact, the world-volume theory we will derive is that on the single wall
(or single antiwall); the presence of the second component is irrelevant.
This is due to our specific choice of relevant parameters.
Neither walls nor strings can be excited in the range of parameters we work.
This means that whatever happens on the surface the
wall, unambiguously fixes what happens on the surface of the
antiwall and {\em vice versa}. In this sense, our holography is peculiar
and is not similar to the situation in string theory.
Our holography relates phenomena seen on one wall (or,
which is the same, on one antiwall)
to the bulk physics of ``minimal" (non-excited, ``straight") strings
stretched between the wall and antiwall.
This is why in fact our world-volume theory has four supercharges,
and the particle supermultiplets on the world volume  are short.

One can try to formalize the above intuitive arguments as follows.
The wall preserves four supercharges determined by the conditions
(\ref{wsusy}).   The antiwall preserves four other supercharges
determined by the same conditions with the opposite sign,
\beqn
\bar{\ve}^2_{\dot{2}} &=& i\ve^{21}\,,\qquad
\bar{\ve}^1_{\dot{2}}=i\ve^{22}\,,
\nonumber\\[3mm]
\bar{\ve}^1_{\dot{1}}&=& -i\ve^{12}\,,\qquad
\bar{\ve}^2_{\dot{1}}=-i\ve^{11}\,.
\label{awsusy}
\eeqn
Let us
consider a new z-dependent SUSY transformation defined on two patches.
The wall patch is at $$\frac{z_1+z_2-L}{2}  < z < \frac{z_1+z_2}{2}\,,$$ while the
antiwall patch is at $$\frac{z_1+z_2}{2} < z < \frac{z_1+z_2+L}{2}\,,$$
where we  denote the wall position by $z_1$ and that
of the  antiwall by $z_2$, assuming
$$l\equiv z_2-z_1 $$
to be large.

Now, define four new
supercharges as selected by the condition (\ref{wsusy}) on the wall patch and by
the  condition (\ref{awsusy}) on the antiwall
patch. This new SUSY transformation acts almost trivially on the
wall-antiwall solution (up to exponentially small terms coming from
the antiwall tail in the wall patch, and the wall tail in the antiwall
patch). As soon as the space is almost empty  at two points
$z=(z_1+z_2)/2$ and $z=(z_1+z_2+L)/2$, there are no problems of
gluing these two patches together (there are  some $\delta$-function
contributions in the SUSY algebra, to be taken care of, but these
 give (almost) zero acting on the wall-antiwall system).
These are exactly the four SUSY charges acting in our world-volume theory.

Let us consider this theory in more detail. The
kinetic terms for the fields $z_1$ and $z_2$ in the world-volume theory are obvious,
(see (\ref{3wwa}))
\beq
 \frac{T_{\rm w}}{2}
\left[
(\partial_n z_1)^2+ (\partial_n z_1)^2\right]
=\frac{1}{2e^2}
\left[(\partial_n a^{(1)}_{2+1})^2 +(\partial_n a^{(2)}_{2+1})^2\right]
,
\label{kinetic}
\eeq
where we use (\ref{az}) to define the fields $a_{2+1}^{(1,2)}$. The sum of these
fields,
$$
a_{+}\equiv \frac{1}{\sqrt{2}}\left(a^{(2)}_{2+1} + a^{(1)}_{2+1}\right),
$$
with the corresponding superpartners, decouples from other fields
forming a free field theory describing dynamics of the center of mass of
our construction. This is a trivial part which will not concern us here.

An interesting part is associated with the field
\beq
a_{-}\equiv \frac{1}{\sqrt{2}}\left(a^{(2)}_{2+1} - a^{(1)}_{2+1}\right).
\label{aminus}
\eeq
The factor $1/\sqrt 2$ ensures that $a_{-}$
has a canonically normalized kinetic term. By definition, it is related to the
relative wall-antiwall
separation, namely,
\beq
a_{-}=\frac{2\pi\xi}{\sqrt{2}}\,l.
\label{al}
\eeq
Needless to say, $a_{-}$ has all necessary ${\cal N}=2$ superpartners.
In the bosonic sector we introduce the gauge field
\beq
A_n^{-}\equiv \frac{1}{\sqrt{2}}\left( A_n^{(1)}-A_n^{(2)}\right),
\label{Aminus}
\eeq
with the canonically normalized kinetic term.

Now,
let us include the classical string solutions attached to the wall and
antiwall. The BPS conditions for the string are given in Eq.~(\ref{ssusy}).
The string-wall-antiwall junction is a 1/4-BPS object.  Two
supercharges acting trivially on this junction are defined by
(\ref{wsusy}) and (\ref{ssusy}) in the wall patch, and by (\ref{awsusy})
and (\ref{ssusy})  in the antiwall patch. Thus,
the string is described  by a short chiral multiplet in the
world-volume theory.

In quantum theory
the strings stretched between
the wall and antiwall, on both sides, will be represented by two chiral
superfields,
$S$ and $\tilde S$, respectively. We will denote the corresponding
bosonic components by $s$ and $\tilde s$.

In terms of these fields the quantum version of the theory (\ref{cl})
is completely determined by the charge assignment (\ref{elch}), (\ref{scch})
and \ntwo super\-symmetry. The charged matter fields
have the opposite electric charges and distinct mass terms, see below.
A mass term for one of them is introduced by virtue of a ``real mass,"
as is explained in Appendix. It is necessary due to the fact that there are
two interwall distances, $l$ and $L-l$.
The real mass breaks parity.
The bosonic part of the action has the form
\beqn
S_{\rm bos} &=&\int d^3 x\,
\left\{-\frac{1}{4e^2}\, F_{mn}^-\,F^{-\,\,mn} +\frac{1}{2e^2}\,
\left(\partial_n\,a_-\right)^2+ \left|  {\cal D}_n s\right|^2 +\left|
\tilde{\cal D}_n \tilde s\right|^2\right.
\nonumber\\[3mm]
&-&
2  a^2_- \, \bar s \,s -2(m-a_-)^2\, \bar{\tilde s} \,\tilde s
-e^2\left(|s|^2-|\tilde{s}|^2\right)^2\Big\}\,.
\label{qu}
\eeqn
We will present additional comments on the derivation of this expression in
Sect.~\ref{physi}.
According to our discussion in Sect.~\ref{vwj},
the fields $s$ and $\tilde s$  have charges +1 and $-1$ with respect to the
gauge fields $A_n^{(1)}$
and $A_n^{(2)}$, respectively. Hence,
\beqn
{\cal D}_n &=& \pt_n -i\left( A_n ^{(1)}-A_n^{(2)}\right)=
\pt_n-i\sqrt{2}A_n^{-}\,,
\nonumber\\[3mm]
\tilde{\cal D}_n
&=&
\pt_n +i\left( A_n^{(1)}-A_n^{(2)}\right)=
\pt_n+i\sqrt{2}A_n^{-}\,.
\label{nabla}
\eeqn
The electric charges of strings with respect to the field $A_n^{-}$ are
$\pm \sqrt{2}$. The last term in (\ref{qu}) is the $D$-term dictated by
supersymmetry.
So far,
$m$ is a free parameter whose relation to $L$ will be determined shortly.
Moreover, $F_{mn}^-= \pt_m \, A_n^ - - \pt_n \, A_m^ -$.
The theory (\ref{qu}) with the pair of chiral multiplets
$S$ and $\tilde{S}$ is free from IR divergences and global $Z_2$ anomalies
\cite{AHISS,BHO}. On the classical level it is clear from our discussion
in the previous section. In Ref. \cite{Tbrane} a version of the
world volume theory (\ref{qu}) with only one supermultiplet
$S$ was considered.

Now we make a crucial test of our theory (\ref{qu}) calculating the masses
of charged matter multiplets $S$ and $\tilde{S}$. From (\ref{qu})
we see that the mass of $S$ is given by
\beq
m_s=\sqrt{2}\,\,\bra a_{-}\ket.
\label{Ms}
\eeq
Substituting here the relation (\ref{al}) we get
\beq
m_s= 2\pi\xi\,l.
\label{Mstr}
\eeq
The mass of the charged matter field $S$ is equal to the mass of the string
of the bulk theory
stretched between the wall and antiwall at separation $l$, see (\ref{tens}).
 Of course, this was
  expected. Note that this is a nontrivial check
of the consistency of our world-volume theory with the bulk theory. Indeed,
 the charges of strings end-points (\ref{elch}) and  (\ref{scch}) are
fixed by the classical solution for the wall-string junction.
%Then the
%coefficient $c$  in
%Eq.~(\ref{qu}) is also fixed.

Now, imposing the relation between the free mass parameter $m$ in (\ref{qu}) and
the length of the compactified $z$-direction $L$ in the form
\beq
m=\frac{2\pi\xi}{\sqrt{2}}\,L
\label{mL}
\eeq
we get the mass of the chiral field $\tilde{S}$ to be
\beq
m_{\tilde{s}}= 2\pi\xi\,(L-l)\, .
\label{Mtstr}
\eeq
The mass of the string $\tilde{S}$ connecting the wall with the  antiwall from
the other side of the cylinder equals string tension times $(L-l)$, in full
accordance with
our expectations, see Fig.~\ref{fig:cylinder}.

To conclude this section let us discuss relations between the  parameters of
the bulk theory we have to impose to ensure that our world-volume theory
(\ref{qu}) makes sense. Most importantly,  we  use the quasiclassical
approximation in our bulk theory (\ref{n2sqed}) to find the solution
for the string-wall junction \cite{SYw} and derive the wall-antiwall world-volume effective theory
(\ref{qu}). This assumes weak coupling in the bulk, $g^2\ll 1$. According to the
duality relation (\ref{duality}) this implies strong coupling in the world-volume theory.

We want to continue the world-volume theory (\ref{qu}) to the
weak coupling regime,
\beq
e^2\ll \frac1{R}\, ,
\label{bwc}
\eeq
which means strong coupling in the bulk theory, $g^2\gg 1$. Our general
idea is that at $g^2\ll 1$ we can use the bulk theory (\ref{n2sqed}) to
describe our wall-antiwall system while at $g^2\gg 1$ we better
use the world-volume theory (\ref{qu}). We call this situation the
{\sl  bulk--brane duality}.
It is quite similar in spirit to the AdS/CFT correspondence.

Besides the condition (\ref{bwc}) or $g^2\gg 1$, we  assume the
condition (\ref{mxi}) to be satisfied too. Now, with the given choice
of the  bulk parameters $\Delta m$, $\xi$ and $g^2$, subject to relations
(\ref{bwc}) and (\ref{mxi}), let us find out whether or not we have a window
of the allowed values of  $L$ (the size of the compactified direction) such
that the theory (\ref{qu}) gives a correct description of
 low-energy world-volume physics.

A lower bound on  $l\sim L$ comes from (\ref{largesep}). It ensures that
the distances between the wall and antiwall from  both sides of the cylinder
are much larger than the wall thickness. An upper bound on $L$ comes
as follows. The string states $S$, $\tilde{S}$ included in the low-energy
action (\ref{qu}) should be much lighter then other massive excitations
on the wall,  with the characteristic mass scale   $\sim 1/R$. Thus, we demand
\beq
m_s\sim m_{\tilde{s}}\ll \frac1{R}\,,
\label{strlR}
\eeq
where the  masses of the string states are given in Eqs.~(\ref{Mstr}) and (\ref{Mtstr}).
This requires, in turn,
\beq
\sqrt{\xi}\ll \frac1{R}\,,
\label{xiR}
\eeq
were we used Eq.~(\ref{largesep}). This is even a more restrictive condition than
Eq.~(\ref{bwc}).

In what follows we will assume that this important condition is satisfied.
Unfortunately, we cannot guarantee it is true. In fact, classically this
would require taking
the coupling of the bulk theory to be
$$g^2\gg \frac{\Delta m}{\sqrt{\xi}}\gg 1\,.$$
Clearly, in this ultra-strong
coupling regime in the bulk our classical expression for $R$ (\ref{R}) is
not valid and
we cannot use it to prove (\ref{xiR}). Below we conjecture that the
condition  (\ref{xiR}) can be reached in the strong coupling regime of the
bulk theory.

It turns out that we have even a more restrictive condition on the
size of the compact direction $L$ than the one in Eq.~(\ref{strlR}).
Indeed, the stretched strings of
the bulk theory are represented in our effective theory on the walls
by two chiral superfields,  $S$ and $\tilde{S}$. This means that the only
properties of the bulk strings seen in the world-volume theory are the
charges of the
string end-points and the string masses. In particular, the walls do not ``feel''
string excitations. The latter would correspond to an  infinite tower
of  extra states in the world-volume theory
with masses $m_{KK}=k/l\sim k/(L-l)$, where $k$ is integer. Let us call these
 states
``KK modes" because to an observer on
the walls they will look similar to  Kaluza--Klein modes. Hence, we require
\beq
m_{KK}\gg m_s\,,
\label{KK}
\eeq
which implies
\beq
L\ll \frac1{\sqrt{\xi}}\,.
\label{upperL}
\eeq
The latter inequality  is equivalent to
\beq
m_{s}\ll \sqrt{\xi}\,.
\label{strxi}
\eeq

\begin{figure}[h]
%\epsfxsize=10cm
%\epsfxsize=6cm
%\epsfysize=2.5cm
%\centerline{\epsfbox{masses.eps}}
 \centerline{\includegraphics[width=4in]{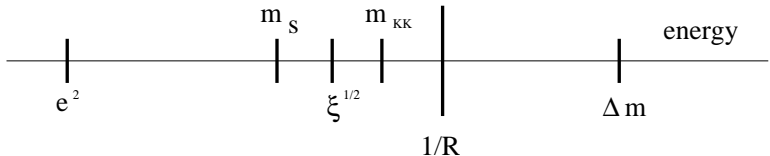}}
 \caption{\footnotesize Mass scales of the bulk and world-volume theories.
}
\label{fig:masses}
\end{figure}

We see that, for the choice of parameters of the bulk theory subject to the
conditions (\ref{mxi}) and (\ref{xiR}), we can assume $L$
to lie inside the window
(\ref{largesep}), (\ref{upperL}) to ensure that our (2+1)-dimensional
theory (\ref{qu}) correctly describes the low-energy limit  of the wall-antiwall world-volume theory, with strings stretched between the
walls fully taken into account. Different scales of our theory are shown in
Fig.~\ref{fig:masses}.

Summarizing, the scales $\Delta m$, $\sqrt{\xi}$ and $e^2_{2+1}\sim \xi/\Delta m$
are determined by the string and wall tensions in our bulk theory, see
(\ref{tens}) and (\ref{wten}).  In particular, the (2+1)-dimensional
coupling $e^2$
is determined by the ratio of the wall tension to the square of the string
tension, as follows from Eqs.~(\ref{3wwa}) and (\ref{az}).
Since the strings and walls in the bulk theory
are BPS-saturated, they receive
no quantum corrections. Equations (\ref{tens}) and  (\ref{wten}) can be continued to
the strong coupling regime in the bulk theory. Therefore, we always can take such values
of
parameters
$\Delta m$ and  $\sqrt{\xi}$  that the conditions
\beq
e^2 \ll \sqrt{\xi}\ll \Delta m
\label{conditions}
\eeq
are satisfied.

To actually prove low energy duality
between the bulk and world-volume theories (\ref{n2sqed}) and (\ref{qu})
 we only need to
prove the condition (\ref{xiR}). This will give us the hierarchy of the
mass scales shown in Fig.~\ref{fig:masses}. With the given values of parameters
$\Delta m$ and  $\sqrt{\xi}$ we have another free parameter of the bulk theory
to ensure (\ref{xiR}), namely, the coupling constant $g^2$.
However, as we explained above, the
scale $1/R$ (the mass scale of various massive excitations living on the wall)
is not protected by supersymmetry and we cannot prove
that the regime (\ref{xiR}) can be reached
at strong coupling in the bulk theory. Thus, our  bulk--brane duality
conjecture is essentially equivalent to the statement that the regime
 (\ref{xiR}) is in fact available under a certain choice of parameters.

Note, that if the condition (\ref{xiR}) is not met, the wall excitations
become  lighter than strings, and the theory (\ref{qu}) does not correctly
describe low-energy physics of the theory on the walls.

\section{Physics of the world-volume theory}
\label{physi}
\setcounter{equation}{0}

\ntwo supersymmetric U(1) gauge theory (\ref{qu}) was studied in
\cite{AHISS,BHO}. It can be obtained by dimensional reduction of \none
QED from four to three dimensions (see Appendix). The neutral scalar field $a_{-}$
comes from the four-dimensional gauge potential upon reduction to three
dimensions.
Terms containing the parameter $m$
 in Eq.~(\ref{qu})  are introduced by virtue of the
real mass procedure \cite{AHISS,BHO}.
To make further consideration more transparent it is instructive to write down here all bilinear fermionic terms,
\beqn
\Delta S_{\rm ferm} &=&\int d^3 x\,
\left\{\frac{1}{e^2}\,\bar\lambda_-
\,i\,\!{\not\!\partial}\,\lambda_- +\bar\psi \,i\,\!{\not\!\!{\cal D}}\,\psi \,
+\bar{\tilde\psi} \,i\,\!{\not\!\!\tilde{\cal D}}\,\tilde\psi \right.
 \nonumber\\[2mm]
&-&
\left.
\sqrt{2}\left( a_{-}\bar{\psi}\psi +
(m-a_{-})\bar{\tilde{\psi}}\tilde{\psi}\right)
\right\}\,,
\label{ferm}
\eeqn
where $\psi^{\alpha}$ and $\tilde{\psi}^{\alpha}$ ($\alpha=1,2$) are the fermion
superpartners of string scalar fields $s$ and $\tilde{s}$.
If the parameter $m$ in (\ref{ferm}) vanished, the masses of the fields
$\psi$ and $\tilde{\psi}$ would be opposite in sign. In this case the theory would be
$P$-invariant (for a definition of $P$-invariance in 3D see Appendix), free of global anomalies and would generate no Chern--Simons
term \cite{Redlich,AlvGauW,AHISS}. However, in our set-up $m$ cannot vanish.
It is related to the size of the compactified dimension $L$ of the
bulk theory, see Eq.~(\ref{mL}). Moreover, since both $l$ and $L-l$ must be
positive, we conclude that $a_{-}$ and $(m-a_{-})$ must be positive too.
In this case $P$-invariance of the world-volume  theory is broken and a Chern--Simons
term is generated.

Let us integrate out the string multiplets $S$ and $\tilde{S}$ and study
the effective theory for the U(1) gauge supermultiplet at scales below
the string masses $m_s$. Once the string fields enter the action quadratically
(if we do not resolve the algebraic equations for the auxiliary fields) the
one-loop approximation is exact.

Integration over the charged matter fields in (\ref{qu}) and (\ref{ferm}) leads to
generation of  the Chern--Simons term with the coefficient
 proportional to
\beq
\frac1{4\pi}\Big[{\rm sign}(a) +{\rm sign}(m-a)\Big]
\varepsilon_{nmk}A^{-}_n \pt_m A_{k}^{-}\,,
\label{cs}
\eeq
see Appendix.
Another effect related to the one in  (\ref{cs}) by supersymmetry is
generation of a nonvanishing $D$-term,
\beq
\frac{D}{2\pi}
\Big[|m-a_{-}|-|a_{-}|\Big]
=\frac{D}{2\pi}(m-2a_{-})\,,
\label{Dterm}
\eeq
where $D$ is the $D$-component of the gauge supermultiplet.
As a result we get from (\ref{qu}) the following low-energy
effective action for the gauge multiplet:
\beqn
S_{2+1}
&=&
\int d^3 x \left\{ \frac{1}{2e^2 (a_{-})}
(\partial_n a_{-})^2-\frac{1}{4 e^2 (a_{-})} ( F_{mn}^{-} )^2
\right.
\nonumber\\[3mm]
&+&\left.
\frac1{2\pi}\varepsilon_{nmk}A^{-}_n \pt_m A_{k}^{-} +
\frac{e^2 (a_{-})}{8\pi^2}
\left(2a_{-}-m\right)^2\right\},
\label{eff}
\eeqn
where we also take into account here a finite renormalization of the bare
coupling constant $e^2$ \cite{IntS,BHOO,BHO}
\beq
\frac1{e^2 (a_{-})}=\frac1{e^2} +\frac1{8\pi |a_{-}|} +
\frac1{8\pi |m-a_{-}|}\,.
\label{rencoup}
\eeq
This is a small effect since $1/e^2$ is the largest parameter (see Fig.~\ref{fig:masses}),
and $a_-$ is stabilized at $m/2$ (see Eq.~(\ref{vac3})).

Note that the coefficient in front of the Chern--Simons term is integer in Eq.~(\ref{eff})
in terms of $k$ (namely, $k=1$; for the definition of the parameter $k$ see
Appendix).
This is because we integrated out two matter multiplets (strings of the
bulk theory). If we had only one matter multiplet, the induced coefficient
in front of the Chern--Simons term would be equal to $k=1/2$, see Appendix.
This would spoil the gauge invariance of the theory and, to compensate for this effect, we
would need to introduce a {\sl bare} Chern--Simons term with half-integer
coefficient. In fact, this is exactly the situation considered in \cite{Tbrane}.
In this paper a single chiral matter multiplet is introduced,
associated with strings attached to the wall only from the one side.
Then the gauge invariance requires introduction of the bare Chern--Simons
term with coefficient $k_0=-1/2$. After integrating out the
matter multiplet,
the effective coefficient in front of the Chern--Simons term is
\beq
k_{\rm eff}= k_0 +\frac12 =0\,.
\label{keff}
\eeq
No net Chern--Simons term is generated in \cite{Tbrane} provided $a_-$ is positive.

In contrast, in the theory we suggest, we have two chiral multiplets
on the wall, describing bulk strings,
$S$ and $\tilde{S}$. Integrating them out produces an integer coefficient
in front of the Chern--Simons term in (\ref{eff}). No bare Chern--Simons
term is required in this case; for a more detailed discussion see  the
end of this section.

The net $P$-invariance violation associated with the
Chern--Simons term in (\ref{eff}) can be seen in the bulk theory.
Three-dimensional $P$-invariance is defined (see Appendix)
as the transformation $x_1\to -x_1$ and $A_1\to - A_1$, while other
coordinates and gauge potentials stay intact. Thus, the direction of the magnetic
flux inside the strings reverses under this transformation. The string
ending on the wall  with the incoming flux goes into
the string ending on the wall  with the outgoing flux, and {\em vice versa}. Clearly, our bulk
configuration breaks this symmetry. This breaking corresponds to
the generation of the Chern--Simons term in the effective theory
on the walls (\ref{eff}).

The most dramatic effect in (\ref{eff}) is the generation of a potential for
the field $a_{-}$ which corresponds to the separation $l$ between the walls.
The vacuum of (\ref{eff}) is located at
\beq
\bra a_-\ket=\frac{m}{2}\,,\qquad l=\frac{L}2\,.
\label{vac3}
\eeq
There are two extra solutions at $a_-=0$ and $a_-=m$,
but they lie outside the limits of applicability of our approach.

\begin{figure}[h]
%\epsfxsize=10cm
%\epsfxsize=6cm
%\epsfysize=2.5cm
%\centerline{\epsfbox{wint.eps}}
\centerline{\includegraphics[width=2.5in]{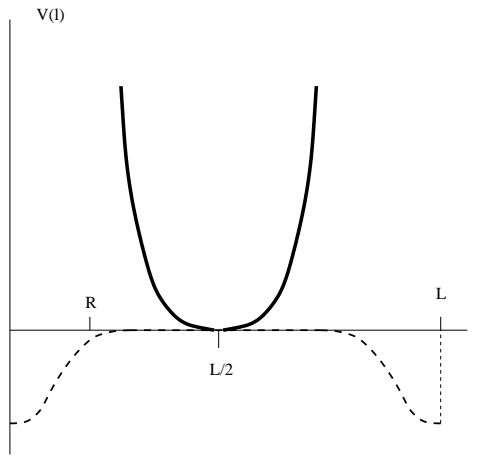}}
 \caption{\footnotesize Classical and quantum wall-antiwall interaction
potential. The dashed line depicts the classical exponentially small potential
while the solid line the quantum potential presented  in Eq.~(\ref{eff}).
Our approximation is not applicable if $l \lsim R$ or $l\gsim L-R$.
It is applicable in the plateau domain which in our approximation is seen as a classically flat direction.}
\label{fig:wint}
\end{figure}

We see that wall and antiwall are pulled apart to be located
at the opposite sides of the
cylinder. Moreover, the potential is quadratically rising with the deviation
from the equilibrium point (\ref{vac3}). As we mentioned in Sect.\ref{quant},
 the wall and anti-wall interact with exponentially small potential
due the overlap of their profiles. However, these interactions
are negligibly small at $l\gg R$ as compared to the interaction in
Eq.~(\ref{eff}). The interaction potential in (\ref{eff}) arises due to
virtual pairs of strings which pull walls together.
Clearly, our description
of strings in the bulk theory was purely classical and we were unable
to see this quantum effect. The classical and quantum interaction potential
of the wall-antiwall system is schematically shown in Fig.~\ref{fig:wint}.
The quantum potential induced by ``virtual strings" is much larger
than the classical exponentially small $W\bar W$ attraction
at separations $l\sim L/2$. It stabilizes the
classically unstable $W\bar W$ system at the equilibrium position (\ref{vac3}).

Note, that if the wall-antiwall interactions were mediated by particles
they would have exponential  fall-off at large separations
$l$ (there are no massless particles in the bulk).
 Quadratically rising potential would never be generated. In our case the
interactions are due to virtual pairs of extended objects -- strings.
Strings are produced as {\sl rigid} objects stretched between walls. We do not
take into account string excitations as they are too heavy, see
Sect.~\ref{quant}. The fact that the strings come out
in our treatment as  rigid objects rather than  local particle-like
states propagating
between walls is of paramount importance. This is the reason why the
wall-antiwall potential does not fall off at large separations.
%Moreover, if an infinite number of string modes had to be taken
%into account, virtual string pair contributions would be suppressed
%exponentially in the coupling constant.
%The suppression we get is just $e^2$.

Note, that power-law interactions between the domain
walls in \none QCD were recently obtained via a two-loop calculation
in the effective world-volume theory \cite{ArmHol}.

\begin{figure}[h]
%\epsfxsize=9cm
%\epsfysize=2.5cm
%\centerline{\epsfbox{syfig3}}
%\centerline{\epsfbox{syfig3.eps}}
 \centerline{\includegraphics[width=2.5in]{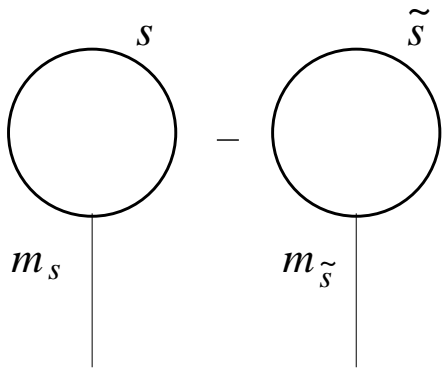}}
 \caption{\footnotesize  Tadpole graphs determining the effective low-energy $D$ term
 in the world-volume theory.}
\label{duf}
\end{figure}

Since the repulsive form of the scalar potential in Eq.~(\ref{eff}),
leading to stabilization at $a_-=m/2$,
is a crucial element of our construction, we would like to add an
alternative argument demonstrating the proportionality
of the effective low-energy $D$ term to $2a_--m$ in the most transparent manner.
Indeed, the above $D$ term can be determined by calculating and summing
two tadpole graphs depicted in Fig.~\ref{duf}.

If $0<a_-<m$, the first graph is proportional to
$m_s= a_-$ while the second one to $m_{\tilde s}=m-a_-$.
The relative minus sign is due to the fact that the electric charges
of $s$ and $\tilde s$ are opposite.

The presence of the potential for the scalar field $a_{-}$ in Eq.~(\ref{eff}) makes this
field massive, with mass
\beq
m_a=\frac{e^2}{\pi}\,.
\label{amass}
\eeq
By supersymmetry, the photon is no longer massless too, it should
acquire the same mass. This is associated with the Chern--Simons term in
(\ref{eff}). As it is clear from the parameter
relations discussed at length in Sect.~\ref{quant} (see also
Fig.~\ref{fig:masses}), $$m_a\ll m_s\,.$$ This shows that integrating out
massive string fields in (\ref{qu}) to get (\ref{eff}) makes sense.

Another effect seen in (\ref{eff}) is the renormalization of the
coupling constant which results in a non-flat metric
of the  target space. Of course, this
effect is very small in our range of parameters since $m_s\gg e^2$.
Still we see that the virtual string pairs induce additional power interactions
between the walls through the nontrivial metric in (\ref{eff}).

To conclude this section, let us note that
our starting point for the quantum theory on the wall world volume
is \ntwo QED, Eq.~(\ref{qu}). In principle, we could add a Chern--Simons
term with arbitrary integer coefficient $k_0$ to our starting theory.
Before excluding the auxiliary field $D$, this amounts to adding
\beq
\Delta S_{\rm CS}=\frac{k_0}{2\pi}\int d^3 x \left\{
\epsilon^{nmk} A_n^-\partial_m\,A_k^-
+ 2(a_- +\mu )D\right\}
\label{CSterm}
\eeq
to the action (\ref{qu}),
where $\mu$  is an  arbitrary parameter of dimension of mass.
Such terms may or may not be present depending on the ultraviolet completion
of the theory. They are not required by the low-energy theory.

We derive our theory (\ref{qu}) using the quasiclassical approximation in
the bulk and, therefore, generally speaking cannot control quantum effects
such as the one in (\ref{CSterm}). One might think
that this term could be induced  in the low-energy theory by some
massive modes living on the wall,  e.g. wall excitations (with mass
$\sim 1/R$), or KK modes mentioned in Sect. \ref{quant}.
However, physics in ultraviolet is determined by the bulk theory which is
$P$-even. Hence, we do not expect a UV-generated Chern--Simons term.
Moreover, if
we formally take the limit $m\to 0$ in our world-volume theory
(\ref{qu}),
we expect that the $P$-invariance should be restored. This rules out
{\em apriori} possible bare Chern--Simons term, so we conclude that
\beq
k_0=0\, .
\label{zerok}
\eeq

\section{Discussion and conclusion}
\label{concl}
\setcounter{equation}{0}

In this paper we presented a field-theoretic system which
possesses a low energy version of
holographic duality. If the bulk 4D world was represented by a cylinder
with a wall and antiwall parallel to each other along the axis of the cylinder
--- just as in the Arkani-Hamed--Dimopoulos--Dvali scenario \cite{ADD} ---
and  were we wall dwellers \cite{dvs},
we would establish that our 3D world is governed (at low energies)
by \ntwo 3D SQED with the Chern--Simons term.
Dynamics of the 4D bulk theory and the bulk set-up on the one hand,
and dynamics of the 3D theory (\ref{qu}) and (\ref{ferm}) on the other hand,
are in one-to-one correspondence. The charged particles
the wall dwellers would discover in their 3D world
would reflect two types of strings (and antistrings)
stretched between the walls.

The holographic description we found is valid in a ``narrow" sense.
Namely, the wall dweller armed with the theory (\ref{qu})  will learn nothing about other excitations in the bulk. Duality that we established is valid
only for the low-energy states. Indeed, e.g. excitations
of the string in the bulk would be represented by an infinite tower of superfields on the wall, which are completely ignored. Excitations of the bulk W bosons are ignored either,
and so are excitations of the walls themselves. This is justified by our choice
of parameters. At the moment we do not know what would happen
if we tried to include (perhaps, some of) higher excitations.
It may well happen that the duality is extendable to a certain extent,
but we do not think it can have the status of the
Kramers--Wannier or sine-Gordon--Thirring or AdS/CFT dualities.

Our construction is self-consistent, both, at the classical and quantum levels.
We start from a classical consideration. In three dimensions \ntwo SQED
with a single matter superfield, in the Coulomb regime, has no finite-energy states.
The theory becomes well defined upon introduction
of the second chiral superfield, with the opposite electric charge.
Correspondingly, the bulk theory must be defined
on the cylinder $R^3\times S^1$, with the $W\bar W$ configuration
parallel to the cylinder axis. Then, the
strings connecting the wall and antiwall
in the bulk, on both sides of the cylinder,
are described in the 3D theory (\ref{qu}) and (\ref{ferm})
by the chiral superfields $S$ and $\tilde S$. Connection on one side of the cylinder
is described by $S$ while on the other side of the cylinder by $\tilde S$.
Besides having the opposite electric charges, the fields $S$ and $\tilde S$
have distinct masses which are proportional to the distance between
$W$ and $\bar W$ on both sides of the cylinder. The sum of the mass terms in
Eq.~(\ref{ferm}) is proportional to cylinder's circumference,
which appears as a ``real mass" in the world-volume theory.

To ensure finiteness of energy at the classical level
one must consider equal number of strings from both sides of the cylinder.
Quantum effects --- virtual string loops ---
generate a mass gap proportional to $e^2$. The fact that this mass gap
depends linearly on $e^2$ is due to infinite rigidity of strings in the approximation
we use. String excitation modes are neglected, which is
perfectly justified under the special choice of parameters we made,
see Sect.~\ref{quant}.

Technically, the mass gap is due to an induced Chern--Simons term.
The origin of this term can be traced back to the cylindrical geometry and the fact
that $L\neq 0$.
This term gives mass to the 3D (dual) photon.
\ntwo supersymmetry imposes the same mass on other members of the vector
supermultiplet. Thus, at the quantum level, the wall and antiwall
start interacting. Their interaction is non-exponential;
rather it depends on the interwall distance as the square of the distance.
The walls are stabilized on the opposite sides of the cylinder.
This is the vacuum state in the world-volume theory.

This feature is absolutely remarkable.
Any interaction between the walls generated by exchange of localized states
(particles) must die off exponentially in the interwall distance.
We would like to explore  in more detail whether power-like dependences
obtained e.g. in \cite{ArmHol} can have a relation to  the phenomenon
observed in this paper.
When our world-volume theory is weakly coupled, if the holographic duality
does indeed take place in our set-up, the bulk theory is strongly coupled.
The impact of the bulk strings at strong coupling is hard to predict.
However, duality does this job for us,
leading to a highly nontrivial prediction of the mass gap generation
(lifting of moduli). In the bulk language, this amounts to a long-range force
between the walls.

It is instructive to discuss how our results
compare with those well-known in string theory
for a system of a $D_p$ brane and anti-$D_p$ brane, see \cite{Sen} for
a review. The instability of the $D_p\, \bar D_p$ system at zero separation
in string theory is associated with the open string tachyon. The tachyon is
the lowest excitation of the fundamental string.
The  $D_p\, \bar D_p$ system eventually collapses: the brane and antibrane
eventually annihilate each other and decay to other objects.

It is not easy to say what plays the role of the
tachyon
in our field-theoretic construction, in which
string excitations are neglected (this is justified by a judicious
choice of parameters) and, therefore, our strings are in essence ``classic."
The fields $S$ and $\tilde{S}$ that represents the ANO strings
on the world volume
are perfectly stable,  with a positive mass squared, see (\ref{Mstr}).
Although our analysis refers to large wall-antiwall separations, we
do not expect any tachyonic behavior of these fields even at
not-so-large values of $l$.

The $a_-$ field of the world-volume theory
is indeed tachyonic outside the fiducial domain.
Inside the fiducial domain its classical  mass squared
is slightly negative, with the exponentially small absolute value.
In the approximation of exact \ntwo on the world volume
 it vanishes, see Fig.~\ref{fig:wint}.

The classical instability of the wall-antiwall
system associated with the exponentially small attraction,
is due to massive $a$ and $q$ fields of the {\sl bulk
theory} rather than  strings.

The difference between our set-up and string-theory  $D_p\, \bar D_p$
system can be, perhaps,
explained as follows. We started from the gauge theory (\ref{n2sqed}).
Therefore, the gauge fields of this theory, as well as
the quark multiplets, are considered as ``fundamental'' fields.
The ANO strings are not fundamental,
they are built of the gauge and quark fields. In contrast, in string theory
the string itself is the  fundamental object. Everything else
is seen as   string excitations. In particular, interactions mediated by
the $a$ and $q$ quanta of the bulk theory
(the reason for the wall-antiwall instability) should be seen
in string theory as tachyons in certain string diagrams.
Note that in string-theory picture the world-volume
gauge fields by themselves are string excitations too, while in our picture
this is not the case, our strings play the role of charged matter for these gauge fields.

In conclusion, let us  point out
an obvious goal for future work: generalization
of the bulk--brane duality studied in this paper to non-Abelian models. To this end, one should consider theories
with non-Abelian domain walls \cite{AbrTow,SYnawall,Jnaw,SakTong}
which, in addition to the domain walls, support non-Abelian strings~\cite{HT1,ABEKY,SYmon,HT2} and 1/4-BPS junctions.

\section*{Acknowledgments}

We are grateful to Arkady Vainshtein for very useful discussions
and to Adi Armoni and David Tong for valuable communications.

The work of M.S.  was
supported in part by DOE grant DE-FG02-94ER408.
The work of A.Y. was  supported
by  FTPI, University of Minnesota, by INTAS grant No. 05-1000008-7865,
RFBR grant No. 06-02-16364a
and by Russian State Grant for
Scientific School RSGSS-11242003.2.

\section*{Appendix: 3D supersymmetric QED}
\renewcommand{\theequation}{A.\arabic{equation}}
\setcounter{equation}{0}

The dual theory on the boundary manifold (the domain walls) is
3D ${\cal N}=2$ supersymmetric QED.
Here we briefly review elements of this theory,
with emphasis on those
which are important in our bulk--brane duality construction.

Let us start from SQED in
four dimensions, with the Fayet--Iliopoulos term $\xi$. The   Lagrangian
of this theory is
\begin{eqnarray}
{\cal L } = \left\{ \frac{1}{4\, e^2}\int\!{\rm d}^2\theta \, W^2 + {\rm
H.c.}\right\} +\sum_f
\int \!{\rm d}^4\theta \,\bar{Q}^f\, e^{c_f\,V}\, Q_f
 - \xi  \int\! {\rm d}^2\theta {\rm d}^2\bar \theta
\,V(\! x,\theta , \bar\theta )
\nonumber\\
\label{sqed}
\end{eqnarray}
where $e$ is the electric coupling constant, $Q_f$ is the chiral matter superfield
(with charge $c_f =
\pm 1$), and
$W_\alpha$ is the supergeneralization of the photon field strength
tensor,
\begin{equation}
{W}_{\alpha} = \frac{1}{8}\;\bar{D}^2\, D_{\alpha } V =
  i\left( \lambda_{\alpha} + i\theta_{\alpha}D - \theta^{\beta}\,
F_{\alpha\beta} -
i\theta^2{\partial}_{\alpha\dot\alpha}\bar{\lambda}^{\dot\alpha}
\right)\, .
\label{sgpfst}
\end{equation}

In  four dimensions the absence of the chiral anomaly in  SQED requires
the matter superfields enter in pairs of the opposite charge. Otherwise the
theory is anomalous, the chiral anomaly renders it
non-invariant under gauge transformations.
Thus, the minimal matter sector includes two chiral superfields
$Q$ and $\tilde Q$, with $c=1$ and $\tilde c=-1$, respectively.
In three dimensions there is no chirality. Therefore, it is totally
legitimate to consider
3D SQED with a single matter superfield $Q$, with  $c=1$.

The 3D theory obtained by dimensional reduction from (\ref{sgpfst})
is well-defined and is $P$-invariant.\footnote{Parity transformation in the
3D theory should be defined as $x_1\to -x_1$ with $x_0$ and $x_2$ intact
\cite{Jackiw}.
If we define gamma matrices in the Minkowskian 3D theory as
$$
\gamma^0 =\sigma_2,\,\,\,\, \gamma^1 =-i \sigma_3,\,\,\,\, \gamma^2 =i \sigma_1
$$
then the action of the $P$ transformation on the 3D fermion field is as
 follows:
$$
\psi\to \sigma_3\,\psi\,.
$$
The kinetic term $i\bar\psi \not\!\!{D}\psi$ is $P$ even, while the mass term
$\bar \psi\psi$ is $P$ odd under this definition.}
This is not the case in the theory with one matter superfield. A remnant of the four-dimensional anomaly in three dimensions
is the so-called parity anomaly. The determinant det$\left(i\bar\psi \not\!\!{D}\psi\right)$
formally is $P$ invariant. However, it is ill-defined
and, in fact, is not gauge invariant \cite{Redlich} (see also \cite{Rdop}).
To make the theory well-defined one needs to add a Chern--Simons term.

Let us start with 3D theory with two oppositely charged matter
supermultiplets obtained by dimensional reduction of (\ref{sgpfst}).
Still we need a certain modification. The theory 
defined by Eq. (\ref{qu}) describes strings on the wall world volume.
In this theory the absolute values of masses of two
oppositely charged chiral multiplets are different.
Such mass terms are well-known, for a review see \cite{3Dzero,BHO,AHISS}.
 They go under the name of ``real masses,"  are specific to theories with
 U(1) symmetries dimensionally reduced from $D=4$ to $D=3$, and present a
direct generalization
of {\em twisted masses} in two dimensions \cite{twisted}. To introduce a
``real mass" one couples matter fields to a background vector field with
a non-vanishing component along the reduced direction. For instance, in the
 case at hand we introduce a background field $V_{\rm b}$ as\,\footnote{Here
the reduced spatial direction is assumed  to lie along the $y$ axis of the
4D space.}
 \beq
 \Delta {\cal L}_m = \int \!{\rm d}^4\theta \,\bar{\tilde{Q}}\,
e^{\,V_{\rm b}}\,  \tilde{Q}\,,\qquad V_{\rm b} =\mu\, (2\,i)
\left(\theta^1\,\bar\theta^{\dot 2}
 -\theta^2\,\bar\theta^{\dot 1}\right).
 \eeq
 We couple $V_{\rm b}$
to the U(1) current of $\tilde Q$ ascribing to $\tilde Q$ charge one with
respect
to the background
field. At the same time $ Q$ is assumed to have $V_{\rm b}$ charge
zero and, thus, has no coupling to $V_{\rm b}$.
Then, the background field generates a real mass term only  for
$\tilde Q$, without breaking ${\cal N}=2$.

After reduction to three dimensions and passing to components (in the Wess--Zumino gauge) we arrive at the action in the following form:
\begin{eqnarray}
S &=&\int d^3 x\,
\left\{-\frac{1}{4e^2}\, F_{\mu\nu}\,F^{\mu\nu} +\frac{1}{2e^2}\,
\left(\partial_\mu\,a\right)^2+\frac{1}{e^2}\,\bar\lambda
\,i\,\!{\not\!\partial}\,\lambda \right.
\nonumber\\[2mm]
&+&\left.
\frac{1}{2e^2}\,D^2  -\xi\,D +\sum_f c_f\bar\phi^f\,\phi_f\, D\right.
\nonumber\\[2mm]
&+&\sum_f\left[
{\cal D}^\mu\bar\phi^f\, {\cal D}_\mu\phi_f
+\bar\psi^f \,i\,\!{\not\!\!{\cal D}}\,\psi_f \right]
-a^2 \bar\phi \,\phi
-(\mu +a)^2 \,\bar{\tilde\phi}\,\tilde\phi
 \nonumber\\[2mm]
&-&  a\, \bar\psi\,\psi +(\mu +a)\, \bar{\tilde\psi}\,\tilde\psi +
% \nonumber\\[2mm]
 %&+&
 \left.\sum_f\, c_f\left[\sqrt{2}\left(\bar\lambda\,\psi_f\right)\bar\phi^f+{\rm h.c.}\right]
\right\}.
\label{bthreed}
\end{eqnarray}
Here $a$ is a real scalar field,
$$
a=-A_2\,,\qquad i{\cal D}_\mu =i \partial_\mu + c_f A_\mu\,,
$$
$\lambda$ is the photino field, and
$\phi_f$ and $\psi_f$ are matter fields belonging to
$Q$ and $\tilde Q$ at $f=1,2$, respectively. Finally, $D$ is
an auxiliary field, the last component of the superfield $V$.

Now let us consider the theory with only one chiral field $Q$. As we will review
below in this case we need to introduce the bare Chern--Simons term
\cite{Redlich,AlvGauW,AHISS}.
In superfields it has the form
\beq
\Delta {\cal L} = c_{\phi}^2\left( \frac{i}{4} \right)
\left( \frac{k_0}{4\pi} \right)\int\,
d^4\theta \, \left[  \bar D^\alpha  V  \right]
\left[  D_\alpha  V  \right]\,,
\label{dvdv}
\eeq
where $c_{\phi}$ is the matter electric charge while $k_0$ is an integer or
half integer number, to be specified below.
In components
\beq
\Delta {\cal L} = c_{\phi}^2\left( \frac{k_0}{4\pi} \right)
\left\{\epsilon^{\alpha\beta\gamma}
\, A_\alpha\, \partial_\beta\, A_\gamma
+2 a\, D +2\,\bar\lambda\,\lambda
\right\}\,.
\eeq

Thus, the bosonic part of  one-matter superfield SQED
takes the form
\beqn
S_{\rm bos} &=&\int d^3 x\,
\left\{-\frac{1}{4e^2}\, F_{\mu\nu}\,F^{\mu\nu} +\frac{1}{2e^2}\,
\left(\partial_\mu\,a\right)^2+ \right.
\nonumber\\[2mm]
&+&
\frac{1}{2e^2}\,D^2  + c_\phi \, \bar\phi \,\phi\, D
+ {\cal D}^\mu\bar\phi \, {\cal D}_\mu\phi
-a^2 \, \bar\phi \,\phi
\nonumber\\[2mm]
&+&\left. c^2_{\phi}
\left( \frac{k_0}{4\pi} \right)\left\{\epsilon^{\alpha\beta\gamma}
\, A_\alpha\, \partial_\beta\, A_\gamma
+2 a\, D \right\}\right\}\,.
\label{sboso}
\eeqn
The fermionic part is
\beqn
S_{\rm ferm} &=&\int d^3 x\,
\left\{\frac{1}{e^2}\,\bar\lambda
\,i\,\!{\not\!\partial}\,\lambda +\bar\psi \,i\,\!{\not\!\!{\cal D}}\,\psi \,- a\, \bar\psi\,\psi \right.
 \nonumber\\[2mm]
&+&
\left.  \left[ \sqrt{2}\left(\bar\lambda\,\psi \right)\bar\phi +{\rm h.c.}\right]
 +
 c^2_{\phi}\left( \frac{k_0}{2\pi} \right)
\,\bar\lambda\,\lambda
\right\}\,.
\label{sfermi}
\end{eqnarray}

Now we can integrate out the
matter multiplet $Q$ assuming that its mass $\bra a \ket$
is large. This will generate an additional Chern--Simons term at the one-loop
level. The effective Chern--Simons coefficient \cite{Redlich,AlvGauW,AHISS}
is given by
\beq
k_{\rm eff}=k_0 +\frac12\,{\rm sign \,}(a)\,.
\label{kren}
\eeq
Gauge invariance requires the coefficient $k_{\rm eff}$ to be
integer. This implies that
the bare Chern--Simons term cannot vanish
in Eqs.~(\ref{sboso}) and  (\ref{sfermi});
rather,  $k_0$ must be
half-integer. This is referred to as the {\sl parity anomaly}. The change of sign
in the fermion determinant under ``large'' gauge transformations is
compensated by nontrivial gauge transformation of the bare
Chern--Simons term with half integer coefficient. Clearly, this problem
does not occur in the theory with two chiral multiplets where
the Chern--Simons term can vanish.

An additional aspect of 3D SQED which we must discuss here is
$C$ parity. Needless to say, the original four-dimensional QED (\ref{sqed})
is $C$-even. The four-dimensional $C$-parity transformation
interchanges
\beq
Q\leftrightarrow \tilde Q\,,\quad A_\mu \leftrightarrow -A_\mu \,,\quad
a \leftrightarrow -a\,.
\label{ddd}
\eeq
This transformation is applicable in the three-dimensional reduced theory
provided no real mass
is added for one of the flavors. Adding such mass we break $Q\leftrightarrow \tilde Q$.
 However, we do need to add real mass.
Moreover, we can  consider 3D SQED with one matter superfield.
The theory will remain  $C$-even with respect to a {\sl different}
$C$-parity transformation,
specific to 3D,
\beq
\psi \leftrightarrow \psi^\dagger\,,\quad \phi \leftrightarrow \phi^\dagger\,,\quad
 A_\mu \leftrightarrow -A_\mu \,,\quad
a \leftrightarrow a\,.
\label{dddd}
\eeq
With respect to this transformation the electromagnetic
charges of the matter quanta $\psi$ and $\phi$ are opposite to those of the antiquanta,
while the scalar charges (i.e. those governing the coupling to the massless field $a$)
are the same for particles and antiparticles of the same flavor.
From the standpoint of four-dimensional $C$ parity, the above three-dimensional
$C$ parity should be viewed as $\left( CP^{(2)}\right)_{D=4}$,
where $P^{(2)}$ stands for reflection of the second axis
(the one which is reduced).

\end{document}